# Spin communication over 30-µm long channels of chemical vapor deposited graphene on SiO$_2$

Z. M. Gebeyehu[a,b,*], S. Parui[c,d], J. F. Sierra[a], M. Timmermans[a], M. J. Esplandiu[a], S. Brems[c], C. Huyghebaert[c], K. Garello[c,*], M. V. Costache[a,*] and S. O. Valenzuela[a,e,*]

[a] Catalan Institute of Nanoscience and Nanotechnology (ICN2), CSIC, The Barcelona Institute of Science and Technology (BIST), Campus UAB, Bellaterra, 08193, Barcelona, Spain.
[b] Universitat Autònoma de Barcelona (UAB), Bellaterra, 08193, Spain.
[c] Imec, 3001 Leuven, Belgium.
[d] K.U. Leuven, 3001 Leuven, Belgium.
[e] Institució Catalana de Recerca i Estudis Avançats (ICREA), 08010 Barcelona, Spain.
*E-mail: zewdu.messele@icn2.cat; kevin.garello@imec.be; marius.costache@icn2.cat; SOV@icrea.cat

*Abstract*

*We demonstrate a high-yield fabrication of non-local spin valve devices with room-temperature spin lifetimes of up to 3 ns and spin relaxation lengths as long as 9 µm in platinum-based chemical vapor deposition (Pt-CVD) synthesized single-layer graphene on SiO$_2$/Si substrates. The spin-lifetime systematically presents a marked minimum at the charge neutrality point, as typically observed in pristine exfoliated graphene. However, by studying the carrier density dependence beyond n ~ 5 x 10$^{12}$ cm$^{-2}$, via electrostatic gating, it is found that the spin lifetime reaches a maximum and then starts decreasing, a behavior that is reminiscent of that predicted when the spin-relaxation is driven by spin-orbit interaction. The spin lifetimes and relaxation lengths compare well with state-of-the-art results using exfoliated graphene on SiO$_2$/Si, being a factor two-to-three larger than the best values reported at room temperature using the same substrate. As a result, the spin signal can be readily measured across 30-µm long graphene channels. These observations indicate that Pt-CVD graphene is a promising material for large-scale spin-based logic-in-memory applications.*

# Introduction

Graphene, a two-dimensional material made up of a one-atom-thick layer of carbon atoms, has spawned great attention because of its unique physical and chemical properties [1], [2]. In particular, graphene has emerged as a tantalizing platform for spin electronics [3]–[5]; owing to its low spin-orbit coupling and lack of hyperfine interaction, spins in it can flow efficiently over very long distances. While exfoliated graphene has been used for fundamental studies, the small scale production limits its suitability for the realization of spintronic technology requiring long propagation channels or multiple devices. This can be overcome by using chemical vapor deposited (CVD) graphene [5]. CVD enables the growth of large area graphene, transfer onto suitable substrates and the fabrication of large-scale spin devices [6], [7]. However, although spin-lifetimes of up to 1.2 ns have been observed in CVD graphene on $SiO_2$, the results are often reported for a single device and are hardly reproducible. This sets serious limitations for testing (complex) circuit architectures with more than one device, which is one of the key requirements for any industrial application. Equally important, the lack of reproducibility precludes, for instance, exploring the spin properties of CVD graphene and the influence of grain boundaries, structural defects and vacancies, which are, for the most part, absent in the exfoliated counterpart. Reliable fabrication of CVD graphene spin devices would, therefore, help enable large-scale spintronic applications and the realization of systematic studies to identify the dominant spin relaxation mechanisms.

In this study, we demonstrate a high-yield fabrication of non-local lateral spin devices with long-distance spin transport at room temperature. The devices are patterned on platinum-based chemical vapor deposition (Pt-CVD) synthesized single-layer graphene on $SiO_2$/Si substrates [8]. We present structural and spin transport characterization of devices with channel lengths $L$ varying from 4 to 30 µm. Spin lifetimes $\tau_s$ of up to 3 ns and spin relaxation lengths $\lambda_s$ as long as ~ 9 µm are observed, which represent the highest values achieved so far for graphene, exfoliated or CVD, on a $SiO_2$/Si substrate. The record-large spin relaxation times and high yield represent a significant advance for spin communication via interconnects and lateral spin-logic technologies [9]. Furthermore, the reproducibility of the measurements allows us to carry out a reliable and exhaustive comparison of the devices characteristics as a function of *L* and graphene carrier density *n* and gather valuable information on the spin relaxation mechanisms. We find that contact-induced spin relaxation does not play a dominant role and that the dependence of $\tau_s$ on *n* agrees with the expectations for spin-relaxation driven by spin-orbit interaction.

## Results

### Device fabrication and characterization

A typical device is shown in **Figure 1a**. It consists of an array of ferromagnetic (FM) contacts, attached to graphene, with variable distances between consecutive electrodes. Pairs of these contacts are used as spin injector and spin detector while their separation defines the spin channel length *L.* We systematically studied the fabrication of the devices to optimize their quality in terms of the graphene source, electron-beam resists, and post-fabrication processing. The raw graphene in the optimized devices was grown on Pt foils at temperatures up to 1100$^o$C by CVD [8]. It was then transferred using electrochemical methods with tetraethyl ammonium hydroxide (0.1M) as electrolyte solution that allow ~100% coverage with good adhesion on SiO$_2$/Si (with 90 nm thick oxide layer) substrates [10], [11]. Large, clean and defect-free areas of graphene were identified and selected using an optical microscope after which long graphene stripes (of about 75 - 150 µm) were lithographically patterned using AR 7520.17 negative-resist based mask and subsequent oxygen-plasma etching. After etching and resist removal, the samples were annealed in a vacuum chamber (10$^{-8}$ Torr) to eliminate resist residues. Ferromagnetic cobalt electrodes were then defined by means of electron-beam lithography using a PMMA/MMA bilayer mask. Titanium oxide barriers were deposited prior to the evaporation of 30-nm thick cobalt to achieve efficient spin injection. Further details on the contact fabrication can be found in Refs. [12] and [13]. The contact arrays have characteristic distances $L$ varying from 4 to 30 μm (**Figure 1a**).

We start by characterizing the structural quality of graphene as well as the electrical properties at room temperature. **Figure 1b** and **c** show, respectively, a typical Raman spectrum and an atomic force microscope (AFM) micrograph of a finished device. The low intensity of the D peak in **Figure 1b** points to a low density of defects in graphene, while the AFM micrograph demonstrates the complete removal of resist residues. In the Raman spectrum, the intensity of the 2D peak is larger than that of the G peak, revealing that graphene is in monolayer form in our devices. Raman measurements at different points of the stripe demonstrate that the graphene has uniform quality across the entire device.

Electrical characterization is carried out by determining the field-effect response with back-gate voltage $V_{gate}$ applied to the (conductive) Si substrate, enabling the control of the graphene carrier density $n$. **Figure 1d** shows the graphene resistance $R$ vs. $V_{gate}$ using four-probe measurements for specific $L$, as defined in a single graphene stripe (see **Figure 1a**). In this stripe, the charge neutrality point (CNP) is found at $V_{CNP}$ = 8 V indicating p-doped characteristics. Although all the devices exhibit p-doping with a spread of just 1 V in $V_{CNP}$ along the stripe, the degree of doping may vary between different batches of samples, with $V_{CNP}$ lying in the range of 8 to 20 V, which implies a residual doping $n_r$ between $2\times10^{12}$ and $5\times10^{12}$ cm$^{-2}$. The field-effect carrier mobility, reproducibly estimated from the slope of the conductivity curve is about μ ~ 1700 cm$^2$V$^{-1}$s$^{-1}$ at $n$ = $5\times10^{12}$ cm$^{-2}$. The contact resistance $R_C$ between the Co-electrodes and graphene, as determined from a three-terminal configuration measurement, is in the range of $R_C$ = 10 to 20 kΩ. As shown in **Figure 1e**, the peak resistance at the CNP, $R_{CNP}$, in four stripes scales linearly with $L$ in the contact array, demonstrating the homogeneity of graphene along the length of the stripe. **Figure 1e** also shows small variations in different regions of the SiO$_2$/Si substrate. Deviation in the slope could arise from a variable density of grain boundaries on different stripes in distant wafer locations [14].

## Spin transport measurements

The spin transport properties are determined using the conventional non-local spin injection/detection scheme [15]. As represented in **Figure 2a**, spin accumulation in the graphene channel is created by injecting a current *I* from a FM tunnel contact (Co2), while a non-local voltage $V_{nl}$ is detected remotely using a second FM contact (Co3) at a distance *L*. A change in the relative magnetization orientation of the injector and detector electrodes, from parallel to antiparallel, leads to a change $\Delta V_{nl}$ in $V_{nl}$. $\Delta V_{nl}$ is a measure of the spin accumulation, or difference in electrochemical potential for spin-up and spin-down carriers, at the detector electrode. The widths of the FM electrodes determine their coercive fields, enabling to switch their magnetizations sequentially by applying an in-plane magnetic field $B_{\parallel}$ along their long axis. The spin signal is characterized by the non-local resistance $R_{nl} = V_{nl}/I$. **Figure 2b** shows typical $R_{nl}$ versus $B_{\parallel}$ in a device with *L* = 20 μm. Spin signals at room temperature in such long graphene channels on any substrate are unprecedented. Previously reported nonlocal signals in graphene on $SiO_2$ were dominated by noise beyond *L* = 16 μm [7], while experiments on exfoliated graphene are typically carried out over shorter channels.

To extract the spin lifetime $\tau_s$ as well as the spin diffusion constant $D_s$ (or the spin relaxation length $\lambda_s = \sqrt{\tau_s D_s}$ ), we perform Hanle spin precession measurements with the magnetic field $B_{\perp}$ applied perpendicular to the graphene plane. In the presence of $B_{\perp}$ the spins undergo Larmor precession while diffusing from the injector to detector. This leads to a modulation of the spin signal, which can be described by the solution of the Bloch diffusion equation:

$$\Delta R_{NL} \propto \int_0^{\infty} \frac{1}{\sqrt{4\pi D_s t}} e^{-\frac{L^2}{4 D_s t}} \cos(\omega_L t) \, e^{-\frac{t}{\tau_s}} dt \qquad \text{Eq. 1}$$

with $\omega_L$ the Larmor frequency.

Figure 2c shows spin precession measurements for parallel and antiparallel magnetization configuration. By fitting the data to Eq. 1 we obtain $\tau_s$ = 3 ns, $D_s$ = 0.03 m$^2$ s$^{-1}$ and $\lambda_s = \sqrt{\tau_s D_s}$ = 9.2 µm.

## Channel-length and carrier-density dependence of the spin signal

The high yield and homogeneity of our Pt-CVD grown graphene-based devices provide the opportunity of studying the spin transport in nearly identical devices as a function of selected parameters, such as the spin channel length *L*, carrier density *n* or tunneling contact resistance. **Figure 3a** shows typical spin precession measurements at $V_{gate}$ = 0 V for *L* ranging from 8 to 27 µm (array 1) with the corresponding fitting curves using Eq. 1. **Figure 3b** presents $\Delta R_{nl}$ as a function of *L*, which yield a linear dependence in a semi-logarithmic representation. **Figure 3c** to **e** show the extracted values for $D_s$ (**Figure 3c**), $\lambda_s$ (**Figure 3d**) and $\tau_s$ (**Figure 3e**) as a function of *L*. The spin parameters are in the range of $\tau_s$ ~ 2.1 - 3 ns, $D_s$ ~ 0.021 - 0.028 m$^2$ s$^{-1}$ and $\lambda_s$ ~ 7 - 9 µm for all the devices, with no particular trend as a function of *L*.

The residual doping and the robustness of the devices warrant the generation of a hole carrier density *n* beyond 10$^{13}$ cm$^{-2}$, which is about a three times larger than typically reached on similar devices [7]. **Figure 4** shows $\Delta R_{nl}$ vs *n* up to $|n|$ ~ 1.1 x 10$^{13}$ cm$^{-2}$. It is observed that $\Delta R_{nl}$ presents a pronounced minimum at the CNP, which is ascribed to a known reduction in $\tau_s$ [16], [17]. However, $\Delta R_{nl}$ reaches a maximum at *n* ~ - 6 x 10$^{12}$ cm$^{-2}$ and then starts decreasing.

In order to indentify the origin of the maximum, we have carried out spin precession measurements as in **Figure 3** using three additional graphene stripes and the corresponding sets of contact arrays. **Figure 5** summarizes $\tau_s$ and $\lambda_s$ as a function of *n* as extracted from such measurements. As discussed previously, all of the devices are *p*-doped, with the doping being homogeneous along each stripe. Both $\tau_s$ and $\lambda_s$ are strongly dependent on *n*. In accordance to observations in ultra-clean graphene devices, originating from exfoliated graphite, $\tau_s$ and $\lambda_s$ present the characteristic minimum about the charge neutrality point (CNP), with $\tau_s$ increasing a factor 2 at $n \sim 10^{12}$ cm$^{-2}$ [16], [17]. However, a remarkable decrease is $\tau_s$ and $\lambda_s$ is found, systematically, for $|n| > 5 \times 10^{12}$ cm$^{-2}$, which has not been previously observed. The decrease is evident in all of the sets of contact arrays.

## Discussion

The spin transport properties in general and the spin lifetime in particular, are determined by the influence of all the relaxation mechanisms available. Comprehensive measurements as those in **Figure 3** and **Figure 5** open alternative routes to determine dominant spin relaxation mechanisms. For example, it has been demonstrated that spin absorption by the contacting electrodes could play an important role in the spin relaxation process when the contacts are transparent [18]. However, it is difficult to assess the contact influence on the spin relaxation in experiments due to strong device-to-device variations. Therefore, it is usually estimated by theoretical modeling [17], [19]–[21].

The results in **Figure 3** allow us to directly address the contact influence on the spin relaxation. As *L* changes, the relative effect of the contacts should follow accordingly. If the contacts dominated the relaxation, their influence would be reflected in the spin transport characteristics. Specifically, as *L* decreases, $\tau_s$ should decrease. However, within our experimental resolution, we do not observe any trend on the spin parameters with *L,* thus we conclude that spin absorption in the electrodes is not a primary cause for spin relaxation in our devices.

The contact influence can be further assessed by plotting $\Delta R_{nl}$ vs *L.* For large enough contact resistances between the FM and graphene, the FMs do not significantly enhance the spin relaxation and $\Delta R_{nl}$ (*L*) is given by $\Delta R_{nl}$ (*L*) $\sim \exp[-L/\lambda_s^*]$, with $\lambda_s^*$ an "effective" spin relaxation length. The consistency of the $\Delta R_{nl}$ measurements in **Figure 3b**, evidenced by the apparent linear dependence, confirms a substantial improvement over prior reports. Here, we consider all of the measured devices in the stripe with *L* ranging from 4 to 27 µm, while equivalent analysis in copper-CVD graphene was limited to $L$ equal to 6 µm or lower, even though measurements were carried out up to $L$ = 16 µm [7].

From the linear fitting (in the semi-log plot) we obtain $\lambda_s^*$ ~ 6.5 µm. This value is somewhat smaller than $\lambda_s$ in **Figure 3d,** as obtained from the Hanle fits. The difference derives from variations in the effective spin polarization $P$ of the injectors and detectors. Indeed, the effective $P$ for the 8-µm channel, $P$ ~ 12%, is unusually high when compared with the estimated value from the other measurements ($P$ ~ 7%). This explains the large spin signal for this specific $L$ value (see **Figure 3b**); if that point was not considered in the fit, $\lambda_s^*$ would be larger than 7 µm. Such observation implies that further enhancement on the device reproducibility should be achieved by improving the tunneling contacts, perhaps using insulating barriers that have shown to be more robust than $TiO_x$, such as $SrO_x$, amorphous carbon interfacial layers or hexagonal boron nitride [17], [22]–[25].

The carrier density dependence of $\tau_s$ and $\lambda_s$ as in **Figure 5** has been analyzed to discriminate between proposed spin relaxation mechanisms in graphene. Initial studies aimed at identifying the relation between spin and momentum scattering times in order to establish whether the spin relaxation responded to an Elliot-Yafet or a Dyakonov-Perel mechanisms, which exhibit opposite scaling [3], [4]. The results were inconclusive or even contradictory. One possible reason is that the scattering events that significantly alter the spin orientation represent only a fraction of all possible scattering events in momentum, rendering a small to null correlation between them. Such would be the case for resonant scattering with local magnetic moments. Calculations on hydrogenated graphene predict spin lifetimes in the range of the experimental results with just ~ 1 p.p.m. of hydrogen [26], [27]. They also show a minimum at the CNP, as observed in experiments. This minimum has also been ascribed to spin relaxation mechanisms involving (Rashba) spin-orbit interaction and the ensued entanglement between spin and pseudo-spin degrees of freedom [28]. Measurements demonstrating isotropic spin lifetime suggest that the spin relaxation is driven by magnetic impurities or random spin-orbit or gauge field [17], [20]. However, the decrease at large $n$ has only been reported for spin relaxation driven by spin-orbit interaction [28], [29] which adds valuable information to solve this puzzle.

## Conclusions

The enhanced room-temperature spin parameters ($\tau_s$ = 3 ns and $\lambda_s$ = 9.2 µm) in our Pt-CVD grown graphene devices represent a step forward in the field of graphene spintronics. These are the largest values reported so far for any kind of graphene on $SiO_2$. Furthermore, the spin signals at ~16 µm are an order of magnitude larger than previously reported [7] while spin information can be transferred over distances of up to 30 µm. The latter represents the longest graphene spin channel to date. Even though larger spin relaxation lengths have been reported in exfoliated graphene encapsulated by, or on, hexagonal boron nitride, the spin channel is much shorter, as it is limited by the exfoliated (random) crystal sizes.

Our devices present unprecedentedly small variability within the same general region of the graphene. An exhaustive investigation of the spin properties as a function of the channel length and carrier density $n$ provides insight into the dominant spin relaxation mechanisms. We find no signs of spin relaxation owing to the contacts and observe, for the first time, a decrease of the spin lifetimes at $n > 5 \times 10^{12}$ $cm^{-2}$. Such behavior has been predicted when the spin-relaxation is driven by spin-orbit interaction and calls for further investigations of the spin lifetime at large $n$ when the spin relaxation is driven by alternative mechanisms, such as resonant magnetic scattering. Beyond fundamental studies, the deterministic high-yield fabrication of our devices is expected to empower investigations of large-scale spintronic applications, such as spin-logic, and logic-in-memory computation for beyond CMOS technologies, which require multiple reliable ferromagnetic contacts and long graphene channels for spin-communication [30],[31].

## Acknowledgment

This research was partially supported by the European Union's Horizon 2020 research and innovation programme under grant agreement No. 696656, by the Spanish Ministry of Economy and Competitiveness, MINECO (under Contracts No. MAT2013-46785-P, No. MAT2016-75952-R, No. MAT2015-68307-P and Severo Ochoa No. SEV- 2017-0706), and by the CERCA Programme and the Secretariat for Universities and Research, Knowledge Department of the Generalitat de Catalunya, 2017 SGR 827. MG acknowledges support from MINECO FPI fellowship under Contract No. BES-2014-069925 and MT from the European Union's Horizon 2020 research and innovation programme under the Marie Sklodoswa-Curie grant agreement Nº 665919.

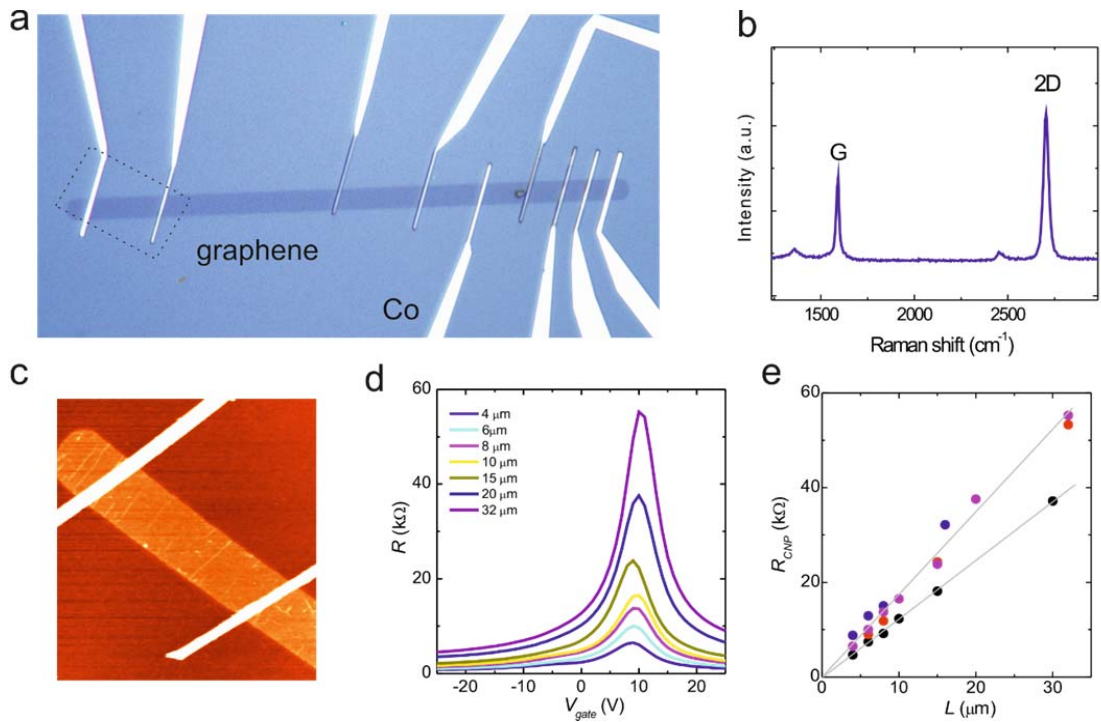

**Figure 1. Characterization of Pt-CVD grown graphene devices**. **a**, Optical image of a 100 μm long graphene stripe and Co-contact array conforming six non-local spin devices with channel lengths varying from 4 to 30 μm. The metal electrodes are made of 30 nm Co on a titanium oxide insulating barriers. **b**, Raman spectrum of the Pt-CVD grown graphene stripe, a.u: arbitrary units. **c**, Atomic Force Microscope micrograph of a region of the graphene stripe including two Co electrodes (first electrodes on the left in **a**); the distance between the electrodes is 13 μm. **d**, Resistance $R$ vs. back-gate voltage $V_{gate}$ measured in a four probe configuration, for different channel lengths $L$. **e**, Resistance at the charge neutrality point ($R_{CNP}$) vs. $L$ in four different contact arrays. The change in the slope from one array to the other might indicate variations in the density of grain boundaries or defects in graphene.

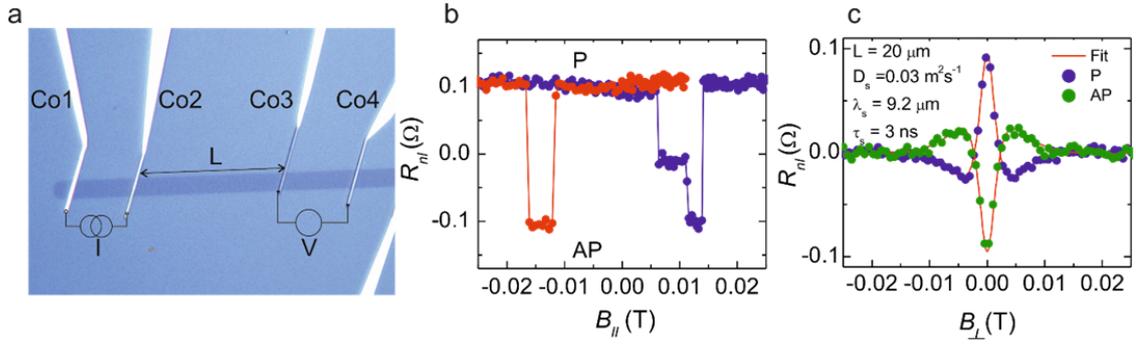

**Figure 2 Non-local spin transport measurements. a,** Optical image of a Pt-CVD grown graphene non-local spin device. For spin transport experiments, the injected current $I$ flows from Co2 to Co1 and the voltage $V_{nl}$ is measured between Co3 and Co4. **b,** Non-local resistance $R_{nl} = V_{nl}/I$ as a function of an in-plane magnetic field $B_\parallel$ applied along the long axis of the ferromagnetic electrodes. **c,** Spin precession measurements with magnetic field $B_\perp$ applied out of the graphene plane for parallel (P) and antiparallel (AP) configuration of the injector/detector magnetizations. The red line represents the fit to the Bloch diffusion equation (Eq. 1). Data in **b** and **c** are acquired in a device with $L = 20$ μm at $V_{gate} = 0$ V.

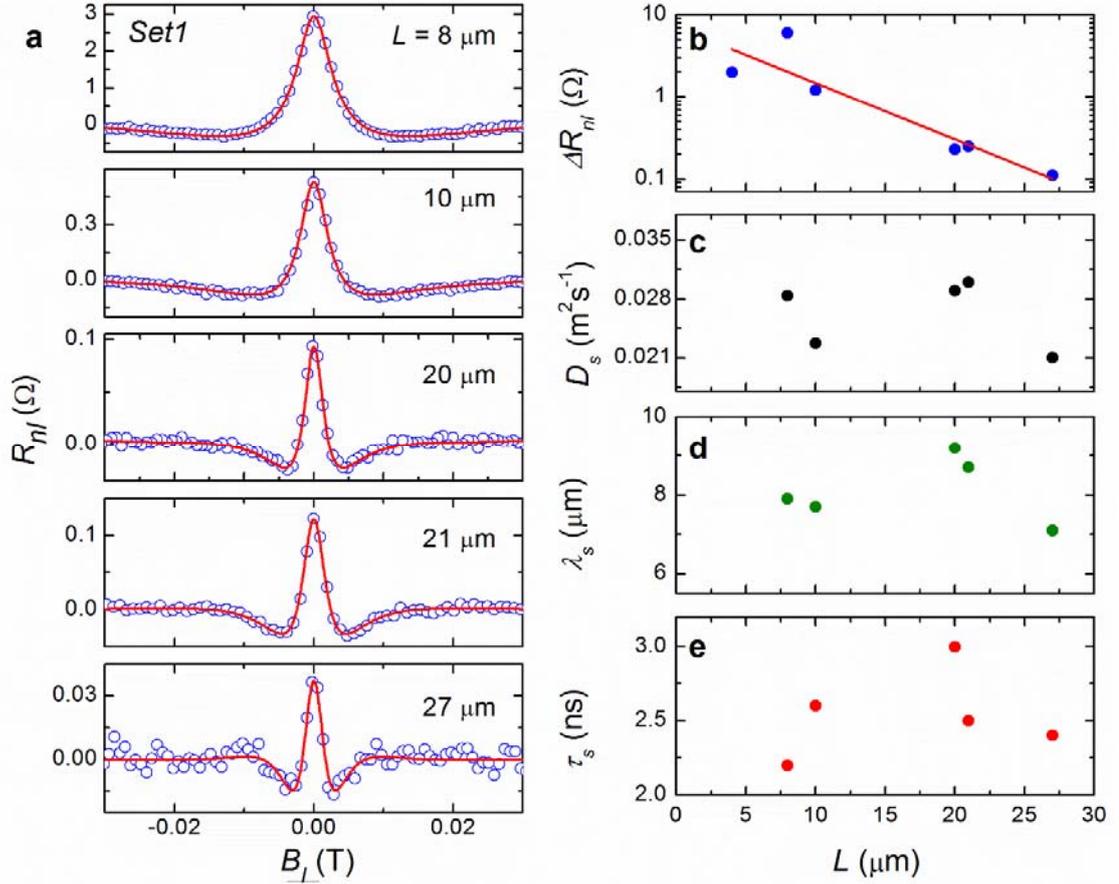

**Figure 3**. **Channel length dependence of the spin transport**. **a,** Spin precession measurements in a graphene stripe, using different pairs of contacts that define the channel lengths, *L*. Open circles correspond to experimental data and solid lines to the associated fit to the solution of the Bloch diffusion equation (Eq. 1.) **b,** Variation of the spin signal $\Delta R_{nl}$ with *L*, the solid line represents an exponential decay with effective spin diffusion length $\lambda^*_s$ = 6.5 ± 1.5 μm. **c,** Spin diffusion constant ($D_s$). **d,** Spin diffusion length $\lambda_s$ and **e,** spin lifetime $\tau_s$ vs. *L* as obtained for the fits in **a**.

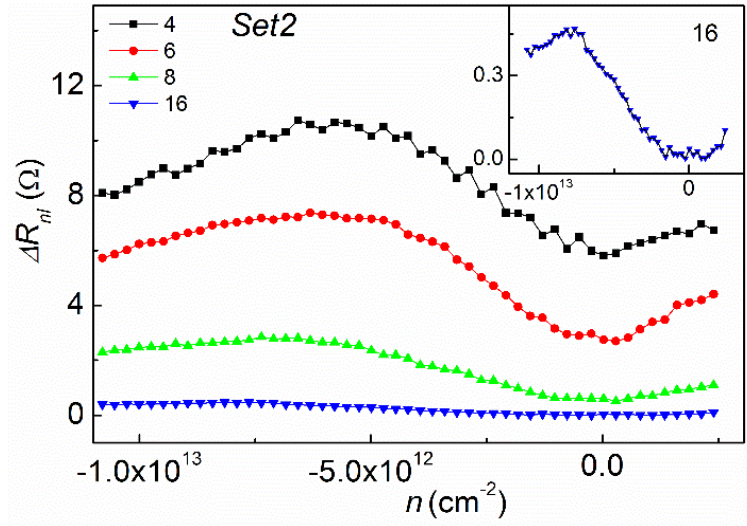

**Figure 4. Spin signal $\Delta R_{nl}$ vs carrier concentration $n$ for different channel lengths $L$.** Regardless of $L$ (shown in μm), $\Delta R_{nl}$ reaches a minimum at the CNP, increases with $n$ up to $n \sim -6 \times 10^{12}$ cm$^{-2}$, beyond which a decrease is observed.

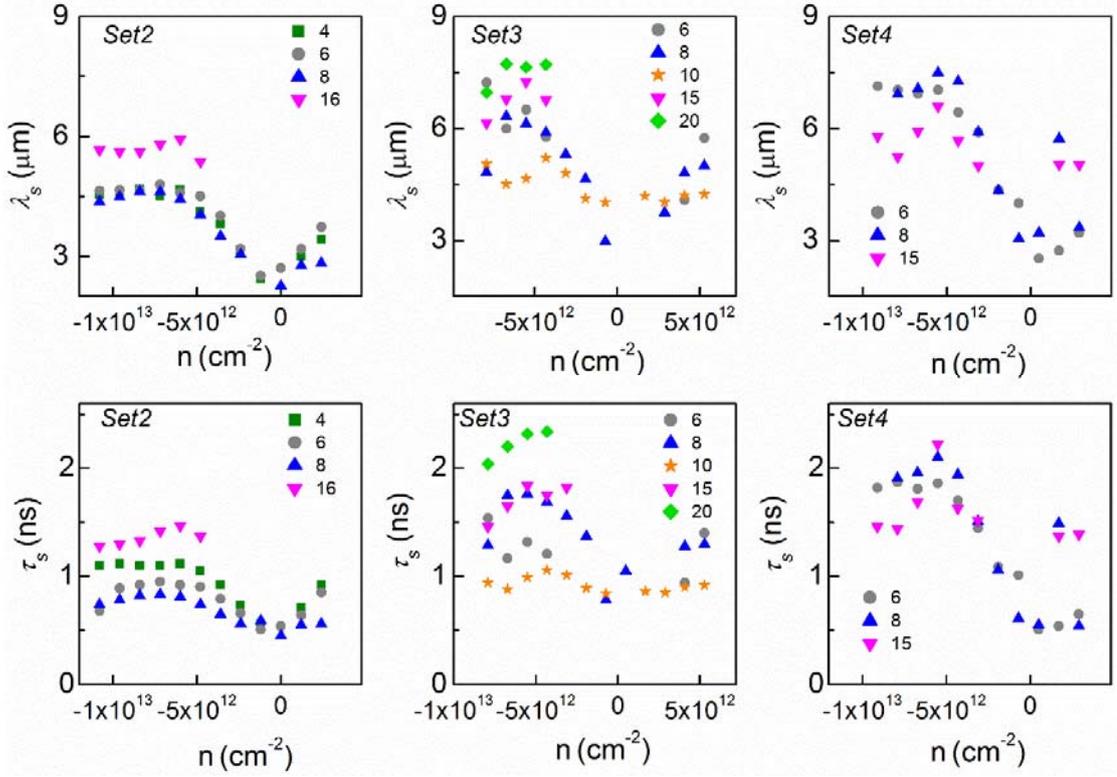

**Figure 5**. **Dependence of the spin properties on three device arrays as a function of carrier density** $n$. The spin diffusion length (top) and the spin lifetime (bottom) are shown for multiple channel lengths $L$. The spin lifetime first increase (up to $n \sim -6 \times 10^{12}$ cm$^{-2}$) then decreases at larger carrier concentrations. The labels indicate $L$ in µm.